\documentclass[traditabstract,preprint]{aa}
\usepackage{epsfig,graphicx,natbib}
\usepackage{longtable}
\usepackage{amssymb}

\def\PKS1830{\hbox{PKS\,1830$-$211}}

\begin{document}

\title{Absolute kinematics of radio-source components in the complete S5 polar cap sample}

\subtitle{IV. Proper motions of the radio cores over a decade and spectral properties}

\author{I. Mart\'i-Vidal\inst{1} \and F. J. Abell\'an\inst{2} \and
     J. M. Marcaide\inst{2} \and J. C. Guirado\inst{2,3} \and M. A. P\'erez-Torres\inst{4} 
     \and E. Ros\inst{5,3,2}  
}

\offprints{I. Mart\'i-Vidal \\ \email{mivan@chalmers.se}}
\institute{Department of Earth and Space Sciences, Chalmers University of Technology, 
           Onsala Space Observatory, SE-43992 Onsala, Sweden  \and
           Departament d'Astronomia i Astrof\'isica, Universitat de Val\`encia,
           C/Dr. Moliner 50, ES-46100 Burjassot, Spain \and
            Observatori Astron\`omic, Universitat de Val\`encia, Parc Cient\'ific, C. Catedr\'atico Jos\'e Beltr\'an 2, 46980 Paterna, Val\`encia, Spain \and
           Instituto de Astrof\'isica de Andaluc\'ia (IAA-CSIC), Apdo. 3004, 18080 Granada, Spain \and 
           Max-Planck-Institut f\"ur Radioastronomie, Auf dem H\"ugel 69, 
           D-53121 Bonn, Germany
}

\date{Accepted for publication}
\titlerunning{Absolute kinematics of the S5 polar cap sample. IV.}
\authorrunning{Mart\'i-Vidal et al. (2015)}

\abstract{

We have carried out a high-precision astrometric analysis of two very-long-baseline-interferometry (VLBI) epochs of observation of the 13 extragalactic radio sources in the complete S5 polar cap sample. The VLBI epochs span a time baseline of ten years and enable us to achieve precisions in the proper motions of the source cores up to a few micro-arcseconds per year. The observations were performed at 14.4\,GHz and 43.1\,GHz, and enable us to estimate the frequency core-shifts in a subset of sources, for which the spectral-index distributions can be computed. We study the source-position stability by analysing the changes in the relative positions of fiducial source points (the jet cores) over a decade.
We find motions of 0.1$-$0.9\,mas among close-by sources between the two epochs, which imply drifts in the jet cores of approximately a few tens of $\mu$as per year. These results have implications for the standard Active Galactic Nucleus (AGN) jet model (where the core locations are supposed to be stable in time). 
For one of our sources, 0615$+$820, the morphological and spectral properties in year 2010, as well as the relative astrometry between years 2000 and 2010, suggest the possibility of either a strong parsec-scale interaction of the AGN jet with the ISM, a gravitational lens with $\sim$1\,mas diameter, or a resolved massive binary black hole.

}


\maketitle

\section{Introduction}

Distant Active Galactic Nuclei (AGN), like quasars and BL Lacs, are currently used as position references in the definition of astronomical inertial frames, from radio \citep[the international celestial reference frame, ICRF][]{Fey,Fey2} to the optical (e.g., {\em Gaia}, \citealt{Linde}). 
The consolidation of reference frames at different regions of the spectrum relies on a well-defined and time-stable chromaticity (i.e., frequency dependence) of the AGN emission. It is well-known that the radio emission from AGN originates at relativistic jets with a frequency-dependent structure. The location of the peak intensity (frequently associated to the so called jet core) depends on the observing frequency due to synchrotron self-absorption effects \citep[e.g., ][]{Blandford,Lobanov}. Due to this effect, the position of the jet core gets closer to the AGN central engine as the observing frequency increases. The study of this self-absorption effect, also known as core-shift \citep[first found by][]{Marcaide89}, has important implications in the study of the jet physics \citep[e.g., ][]{Lobanov}, but it is also crucial for a proper panchromatic alignment of the AGN-based inertial reference frames \citep[e.g., ][]{Kovalev}.

The sky location of AGN cores may not only depend on frequency, but also on time. If the opacity in the jet changes (owing to variability in the particle density and/or the magnetic-field structure) or the jet changes its orientation (e.g., owing to precession), the position of the core at any given frequency (and also the core-shift) evolves. This kind of an evolution of core positions encode information about the changing physical conditions at the innermost regions of the AGN jets, and also map into time-dependent misalignments among AGN-based reference frames at different frequencies. 

To date, a large fraction of geodetic and astrometric very-long-baseline-interferometry (VLBI) observations rely on the group-delay observable. The group-delay astrometry does not usually take the effect of source structures into consideration, whose time variability (and frequency dependence) can introduce astrometric biases of even several times the nominal astrometric group-delay precision \citep{Moor}.
Restricting the observations to very compact jet structures \citep[i.e., jets with low ``structure indices''][]{Charlot} and/or to jets with smooth profiles in a particle-field energy equipartition, help us to minimize the frequency (and time) astrometry variations in the definition of the reference frames with group-delay astrometry \citep{Porcas}. 
But, in any case, the use of phase delays instead of group delays provides a better solution for accounting for the source structure in the astrometry. 
Moreover, the phase delays are more precise than the group delays by up to several orders of magnitude \citep[see][for a deeper discussion]{PaperIII,Tesis}. 

In recent decades, we carried out a set of very-long-baseline-array (VLBA) observations of the complete S5 polar cap sample \citep{Eckart} at 8.4, 15, and 43\,GHz \citep[][hereafter, Papers I and II, respectively]{PaperI, PaperII}. The S5 polar cap sample consists of 13 radio-loud AGN that are located at high declinations (circumpolar for the VLBA). The main goals of this campaign were the study of the frequency dependence and time stability of the jet structures (especially, the jet cores), as well as the characterization of the absolute kinematics of the optically-thin jet components of all sources. All observations were performed in phase-referencing mode, to enable us the use of differential phase-delays in the astrometry analysis of the source positions. The differential phase-delays are the most precise interferometric observables and encode robust information on the relative position of the sources \citep[e.g., ][]{Marcaide89}.
We have published partial results about the evolving source structures at 8.4 and 15\,GHz (Paper I/II), as well as the first astrometry analysis at 15\,GHz, together with a description of our astrometry technique \citep[][hereafter Paper III]{PaperIII}. 
Astrometry results on small subsets of this source sample have also been reported \citep[e.g., ][]{P01,P00}

This is the fourth paper in this publication series. 
In this paper, we report new results from the latest observations of this campaign, which were performed in year 2010 at two frequencies, 14.4 and 43.1\,GHz, using, for the first time in this project, the fast frequency-switching (FFS) observing capabilities of the VLBA \citep[see e.g., ][]{Middelberg}.  
In the next section, we describe our observations. In Sect. \ref{calibration}, we describe the calibration strategy. In Sects. \ref{results15} and \ref{results43}, we present our results at 14.4\,GHz and 43.1\,GHz, respectively. In Sect. \ref{results1543}, we compare the observations at both frequencies and present spectral-index images for a subset of sources. In Sect. \ref{conclusions}, we summarize our conclusions.

\section{Observations}
\label{observations}

The VLBA observations were performed in 2010 December 18, starting at 01:26 (UT) with a duration of about 24 hours. The recording rate was set to 256 Mbps and the observations were in single-polarization mode (only the left circular-hand polarization, LCP, was registered). We covered a total bandwidth of 64\,MHz, divided into eight intermediate frequency bands (IFs). We used the fast frequency-switching (FFS) capabilities of the VLBA frontends, which enabled us to change among different observing bands in approximately half a minute, without loss of coherence among band switches \citep{Middelberg}. Our lowest reference frequency was 14.35099\,GHz (hereafter 14.4\,GHz) and the highest reference frequency was 43.10099\,GHz (hereafter 43.1\,GHz) Thus, the higher frequency was very nearly three times (3.003346) the lowest frequency. Such a frequency configuration makes it possible to perform an intra-source dual-frequency calibration, also called source-frequency phase referencing, \citep{Middelberg, Rioja} and thus help us to determine robustly the core-shifts of all sources between these two observing frequencies.

In Table \ref{SOURCELIST}, we list all the sources of the S5 polar cap sample, together with the short aliases used in this publication (the same as in Paper III) and the coordinates estimated in Paper III. We take the positions reported in Paper III (i.e., those estimated at 15\,GHz in year 2000) as the initial positions for the fit of the 2010 observations. Hence, any shift observed from the observations reported here can be directly related to a physical shift in the source positions between June 2000 (i.e., the epoch reported in Paper III) and December 2010 (i.e., our new observations).

\begin{table}
\caption{Individual sources observed. The positions are those determined in June 2000 \citep{PaperIII}.}
\label{SOURCELIST}
\begin{center}
\begin{tabular}{ccrr}\hline\hline
Source name & Alias & \multicolumn{1}{c}{Right Ascension} & \multicolumn{1}{c}{Declination} \\
            &       & \multicolumn{1}{c}{J2000}   & \multicolumn{1}{c}{J2000}  \\
\hline       
B0016+731 & 00 & 00$^h$ 19$^m$ 45.7862$^s$ & 73$^\circ$ 27' 30.0167'' \\
B0153+744 & 01 & 01$^h$ 57$^m$ 34.9649$^s$ & 74$^\circ$ 42' 43.2289'' \\
B0212+735 & 02 & 02$^h$ 17$^m$ 30.8132$^s$ & 73$^\circ$ 49' 32.6213'' \\
B0454+844 & 04 & 05$^h$ 08$^m$ 42.3635$^s$ & 84$^\circ$ 32' 04.5440'' \\
B0615+820 & 06 & 06$^h$ 26$^m$ 03.0062$^s$ & 82$^\circ$ 02' 25.5678'' \\
B0716+714 & 07 & 07$^h$ 21$^m$ 53.4485$^s$ & 71$^\circ$ 20' 36.3630'' \\
B0836+710 & 08 & 08$^h$ 41$^m$ 24.3653$^s$ & 70$^\circ$ 53' 42.1724'' \\
B1039+811 & 10 & 10$^h$ 44$^m$ 23.0628$^s$ & 80$^\circ$ 54' 39.4428'' \\
B1150+812 & 11 & 11$^h$ 53$^m$ 12.4991$^s$ & 80$^\circ$ 58' 29.1536'' \\
B1749+701 & 17 & 17$^h$ 48$^m$ 32.8403$^s$ & 70$^\circ$ 05' 50.7687'' \\
B1803+784 & 18 & 18$^h$ 00$^m$ 45.6840$^s$ & 78$^\circ$ 28' 04.0183'' \\
B1928+738 & 19 & 19$^h$ 27$^m$ 48.4952$^s$ & 73$^\circ$ 58' 01.5698'' \\
B2007+777 & 20 & 20$^h$ 05$^m$ 30.9987$^s$ & 77$^\circ$ 52' 43.2471'' \\
\end{tabular}
\end{center}
\end{table}

\begin{figure*}[th!]
\centering
\includegraphics[height=5.75cm]{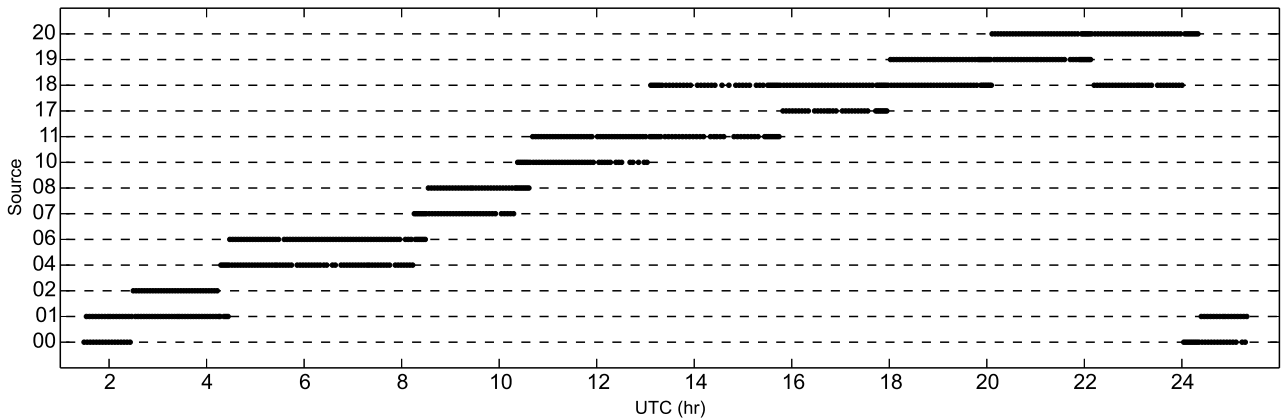}
\caption{Time distribution of our observations on year 2010. Sources under similar time windows were observed in common duty cycles. The pair 18$-$19 was not observed in year 2000.}
\label{SCHEDULE}
\end{figure*}

The observations were arranged in duty cycles. Each duty cycle covered a subset of two to four close-by sources (with integration times between 30 and 60 seconds in each source pointing), in a similar way as the duty cycles described in Paper III. The duty cycles were designed to maximize the antenna elevations, which optimize the quality of the differential phase delays. We show in Fig.\,\ref{SCHEDULE} the time distribution of all the observations. The frequency-switching changes were applied in two types of duty cycles, which were alternated every four iterations. In the first type of cycle (optimized for the frequency-switching calibration, see Sect. \ref{FFSCal}), half of the switchings were applied while the antennas were slewing among sources. If $A_L$ and $A_H$ are observations of source $A$ at 14.4\,GHz and 43.1\,GHz, respectively, the duty cycles were arranged as

$$ A_L - B_H - B_L - A_H - A_L - B_H - B_L...$$

This approach saves some time due to switching. In the second type of duty cycle, observations of different sources at the highest frequencies were put close in time: 

$$ A_L - A_H - B_H - B_L - B_H - A_H - A_L...$$

This approach minimizes the time lag among consecutive observations of different sources at the highest frequency band (43.1\,GHz), where the atmospheric effects are more critical for the phase connection of the differential phase delays. Due to the dual-frequency observations, the duty cycles were, on average, longer in time than those of the epoch of year 2000 (reported in Paper III). Hence, we restricted our duty cycles to close-by sources, to minimize the slewing time and ensure a successful phase connection. The source pairs observed in the duty cycles that have been used in this analysis are listed in Table \ref{CSlist}.

The data were correlated at the National Radio Astronomy Observatory (NRAO) headquarters (Socorro), using the NRAO version of the DiFX software correlator \citep[][]{Deller}. A total of 128 spectral channels per visibility were generated (16 channels per IF). 

\begin{table}
\caption{Source pairs observed. The separations, used as \textit{a priori} in this work, are those determined in June 2000 \citep{PaperIII} using source 07 as reference (see text). }
\label{CSlist}
\begin{center}
\begin{tabular}{lr}\hline\hline
Pair & Separation (deg) \\\hline
01 - 00 &   6.7707315679 \\
01 - 02 &   1.6149364850 \\
04 - 06 &   3.3327854279 \\
08 - 07 &   6.4191401883 \\
11 - 10 &   2.6993766419 \\
11 - 18 &  14.8392077764 \\
18 - 17 &   8.4082929389 \\
18 - 20 &   6.3423054016 \\
19 - 20 &   4.5218913640 \\
\end{tabular}
\end{center}
\end{table}

\section{Data reduction and calibration}
\label{calibration}

\subsection{Fringe finding and phase connection}
\label{calibr}

The calibration was performed using the astronomical image processing system (AIPS) software\footnote{\texttt{http://www.aips.nrao.edu}} by NRAO, using standard procedures, as there are described in Paper III. The dispersive (i.e., ionospheric) delay contribution was removed with the AIPS task TECOR, using GPS satellite data as described in Paper III. The effect of source structures at each band was removed by obtaining CLEAN hybrid images of all sources and using the resulting source models in the computation of the model phases, prior to the final fringe search. The positions of the peak intensities of all the sources at each band were used as the source phase centers (i.e., the fiducial reference points for the astrometry). The total (phase and group) delays were then exported from AIPS for their later analysis with our astrometry software, the University-of-Valencia Precision Astrometry Package \citep[UVPAP, Paper III; ][]{Tesis}. See Paper III for more information about the calibration and the analysis procedure.

The group delays at 14.4\,GHz of all sources were used to derive good a-priori models for the atmospheric non-dispersive delay and the drifts of the stations clocks. These models were then used to perform a preliminary connection of the (otherwise $2\pi$ ambiguous) phase delays. The remaining unmodelled phase cycles were derived using an automatic phase-connection algorithm \citep[][and Paper III]{Tesis}. 

With the phase-delay ambiguities properly corrected at 14.4\,GHz, we finally computed the differential phase-delays (i.e., differences among delays for sources observed in the same duty cycles).  
Typically, the inclusion of the differential delays in the astrometric analysis improves the precision by roughly an order of magnitude, when compared to an ordinary phase-referencing analysis. This is due to the many redundancies present in our multi-source duty-cycle scheduling, and to the superior quality of a parametric astrometry analysis (i.e., fitting delays and phases to a complete geodetic + astrometric model) when compared to ordinary phase-referencing astrometry (where the geodetic + atmospheric models cannot be optimized). See Paper III and \cite{Tesis} for a more complete comparative discussion about the astrometric precision with differential phase-delays. 

\subsection{Frequency-switching phase-transfer calibration}
\label{FFSCal}

The phase connection at 43.1\,GHz is especially difficult, since the delay corresponding to one $2\pi$ phase cycle is so short (only about 23\,ps) that very small unmodelled atmospheric effects can add several $2\pi$ cycles to the phase delays between two consecutive observations of the same source. For this reason, and because there is an ongoing parallel effort in this direction as part of a doctoral thesis, we have not attempted to connect the phases at 43.1\,GHz.

The observations reported in this paper were performed using the FFS capabilities of the VLBA. These capabilities enabled us to make a phase transfer between 14.4\,GHz and 43.1\,GHz using an adaptation of the  
source-frequency phase-referencing (SFPR) method described in \cite{Rioja} \citep[see also][]{Rioja2}. The phase-transfer calibration was performed using an in-house developed software, which makes use of the scriptable \texttt{ParselTongue} interface to AIPS \citep{Parsel}.
 
\begin{figure*}[th!]
\centering
\includegraphics[width=12cm]{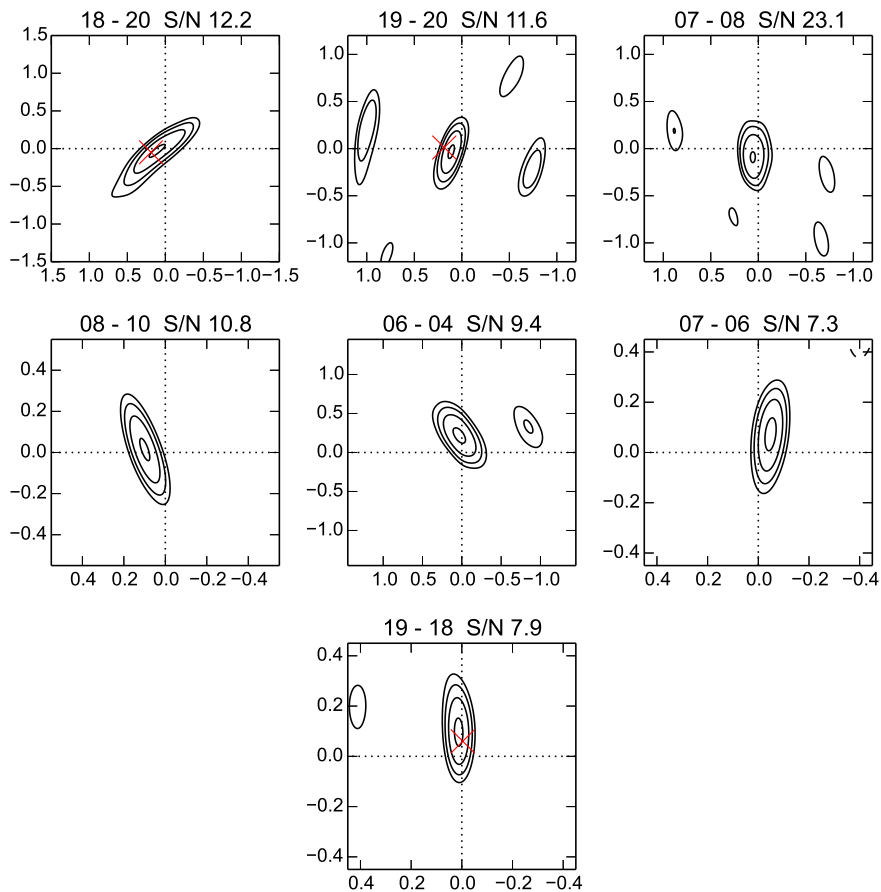}
\caption{ SFPR images at 43\,GHz. The axes, given in mas, correspond to the RA and Dec offsets with respect to the peaks at 14.4\,GHz. Contours are logarithmically spaced between 5$\sigma$ and the source peak. The label XX-YY indicates source YY phase-referenced to the calibrator XX. The red crosses indicate the expected peak positions of sources 18, 19 and 20, assuming that their compact optically-thin jet components (as well as those of their calibrators) are co-spatial at 14.4 and 43.1\,GHz.}
\label{SFPR_IM}
\end{figure*}

\begin{figure*}[th!]
\centering
\includegraphics[height=5.75cm]{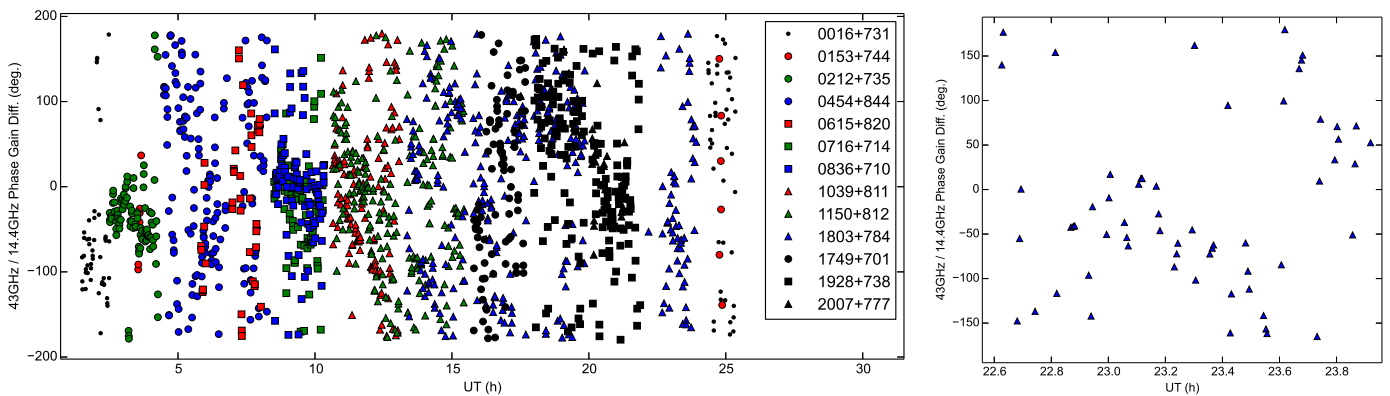}
\caption{Left, difference between the phase gains at 43.1\,GHz and the --scaled up by a factor of three-- phase gains at 14.4\,GHz  for the Pie Town antenna. The instrumental phase offset has been removed (see text). Hence, the differences among sources observed in common duty cycles encode information related to the core-shift of the sources between the two observing frequencies. Right, zoom for source 1803$+$784.}
\label{GAINCOMP}
\end{figure*}

Once we accounted for the source structures in the fringe fitting, the remaining antenna gains at our two observing frequencies were only affected by atmospheric, instrumental and chromatic effects (e.g., core-shifts) in the source structure. The bulk of the ionospheric contribution was removed using the AIPS task TECOR. The non-dispersive contributions were removed by scaling the phase-like antenna gains at 14.4\,GHz by the frequency ratio  ($43.1/14.4\sim3$), to calibrate the 43.1\,GHz data.
Since the observing times at 14.4\,GHz and 43.1\,GHz do not coincide (there is a difference between consecutive scans of at least 30\,seconds, which is the time needed by the FFS system to switch between observing bands), we had to interpolate the gains at 14.4\,GHz to the observing times at 43.1\,GHz, using the rate integral at 14.4\,GHz and accounting for the phase ambiguities among the consecutive 14.4\,GHz observations. For each scan, the instrumental and ionospheric offsets between the scaled 14.4\,GHz phases and the 43.1\,GHz phases were subtracted by phase-referencing from sources observed in common duty cycles (see Table \ref{SOURCELIST2}). After applying the SFPR calibration, we measured the position shifts of the intensity peaks in all the resulting images (shown in Fig.\,\ref{SFPR_IM}). These shifts contain the core-shift of the target sources plus the core-shifts of their respective phase-referencing calibrators (Table \ref{SOURCELIST2}). In order to decouple the shifts of the calibrators from those of their targets, we re-referenced the shifts in all the SFPR images to common points on the sky: the compact and optically-thin jet components in sources 10, 11, 18, 19, and 20 (shown in Fig. \ref{CSFig}). In Fig.\,\ref{SFPR_IM}, we also show the expected SFPR peak positions for the pairs 19-20, 19-18 and 18-20, by assuming that the compact optically-thin jet features at 14.4 and 43.1\,GHz are co-spatial. These results cannot be directly compared to those in \cite{Rioja2015}, since the spatial resolutions are very different and jet-blending effects introduce additional shifts of the source peaks at each frequency. The strategy of using compact optically-thin features as an astrometry reference has been applied in previous core-shift studies \citep[e.g.,][]{Kovalev,Fromm}. We notice that the use of more extended optically-thin emission (e.g., the jet extension in source 08) may bias the core-shift, due to spectral gradients accross the jet structure. In Fig. \ref{GAINCOMP}, we show the difference between the overall Pie Town phase gains at 14.4\,GHz (scaled up by a factor of three, to convert them into 43.1\,GHz gains, and interpolated in time) and the overall phase gains computed directly from the  43.1\,GHz fringes. We notice that the differences in the phase gains for most of the sources are not random. This is indicative of a successful phase-transfer calibration. However, there were a few successful gain solutions at 43.1\,GHz for source 01, which made it not possible to perform the SFPR calibration among sources 00, 01, and 02. In addition, we only considered as successful SFPR detections those with an image dynamic range S/N $> 5$. As a consequence, the core-shift of sources 11 and 17 could not be re-referenced to any other source of the sample.

\section{Results at 14.4\,GHz}
\label{results15}

\subsection{Source structures}

\begin{figure*}[th!]
\centering
\includegraphics[width=12cm]{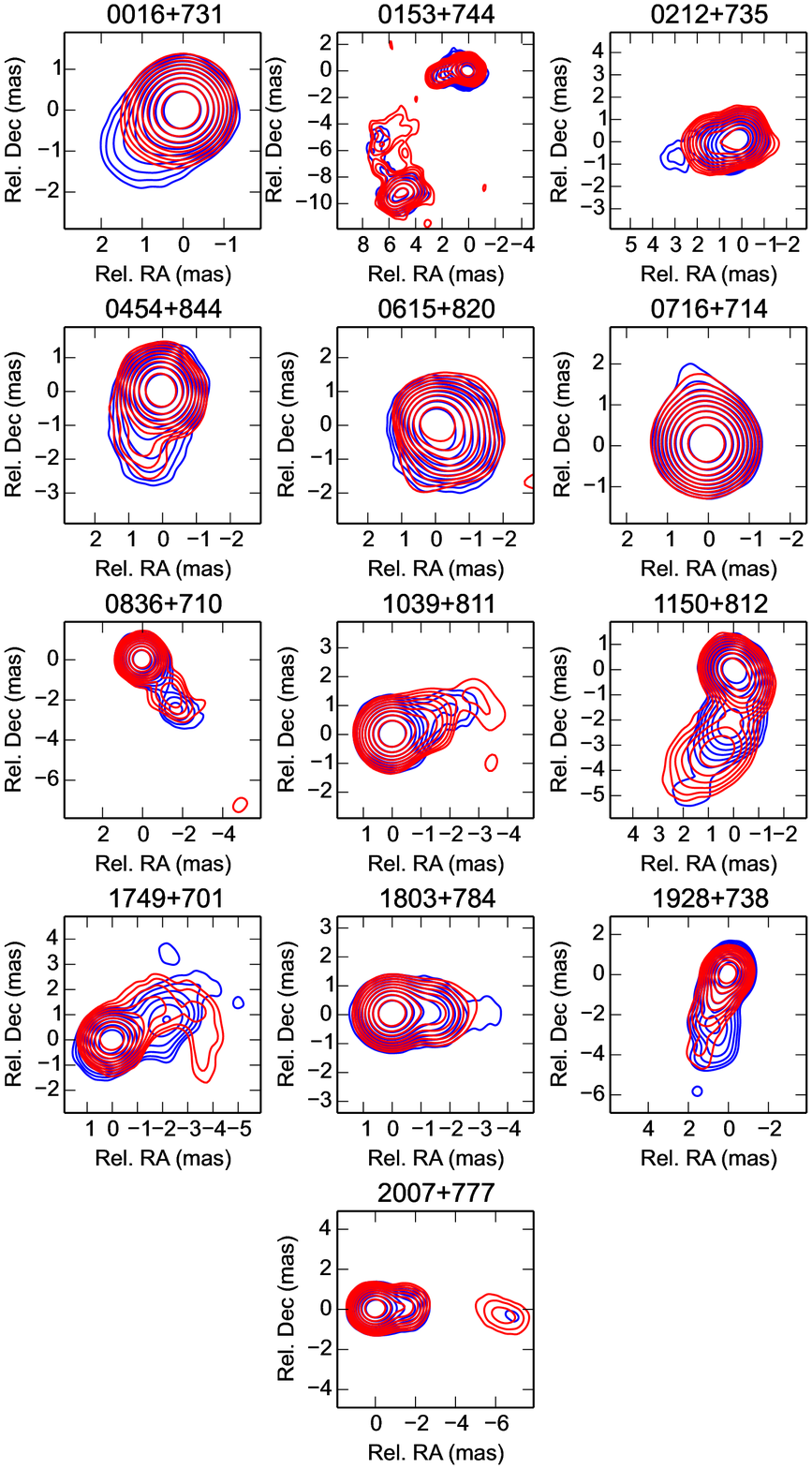}
\caption{Images of the S5 polar cap sample sources in year 2000 (at 15\,GHz, in blue contours) and in year 2010 (at 14.4\,GHz, in red contours). The sources have been shifted to set their intensity peaks at the coordinate origin. The contours are spaced logarithmically from 0.75\% to 99\% of the source peaks. The restoring beams have a FWHM of 1$\times$1\,mas.}
\label{sources15}
\end{figure*}

We show the source structures of all sources of the S5 polar cap sample at 14.4\,GHz in Fig. \ref{sources15}. The structures in year 2000 (i.e., the observations reported in Paper III) are shown in blue contours, and the structures recovered in 2010 in red contours. All sources have been shifted to set their intensity peaks (i.e., the phase centers in our astrometry analysis) at the coordinate origin of each image. The ten contours shown are spaced logarithmically, from 0.75\% to 99\% of the source intensity peaks. The restoring beams have a full width at half maximum (FWHM) of 1$\times$1\,mas in all cases (this is close to the typical major axis of the restoring beams in all sources, using natural visibility weighting). 

\subsection{Differential phase-delay astrometry}

\begin{figure*}[th!]
\centering
\includegraphics[width=18cm]{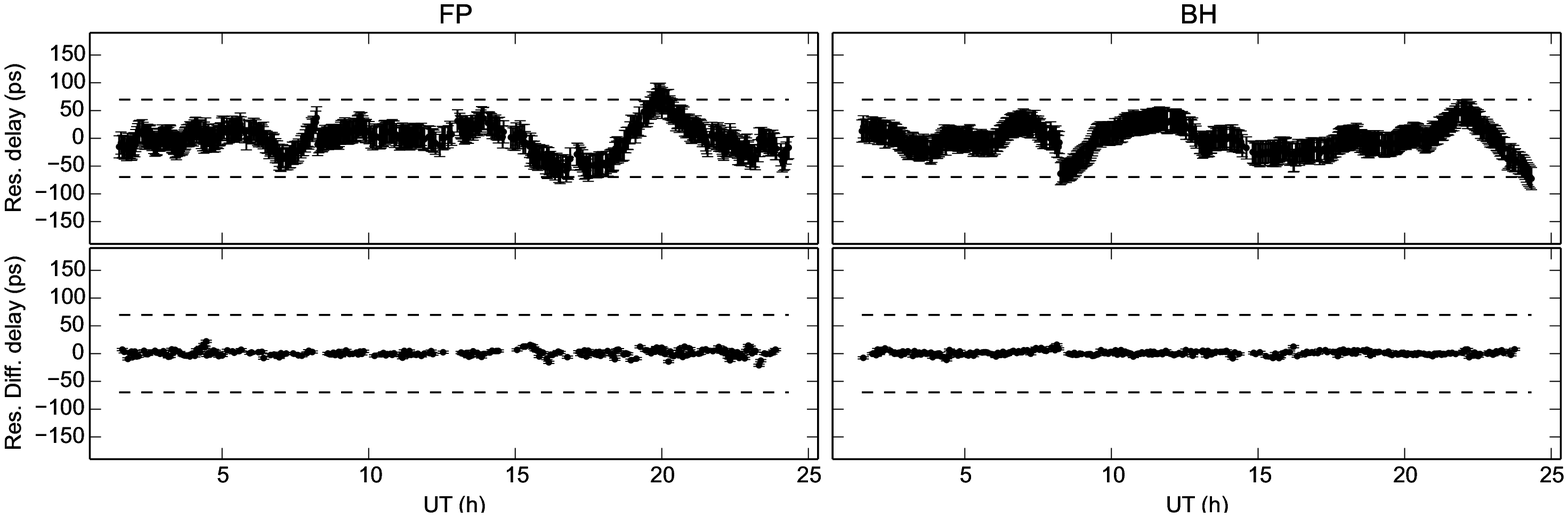}
\caption{Post-fit residual phase delays for Fort Davis -- Pie Town (left) and Brewster -- Hancock (right) for all observed sources. The error bars are shown in all figures: undifferenced delays (top);  differenced delays (bottom). The dashed lines correspond to the delays of a $\pm 2\pi$ phase ambiguity.}
\label{residuals}
\end{figure*}

In Fig. \ref{residuals}, we show the undifferenced and differenced phase delays for two representative baselines (Fort Davis to Pie Town, FP, and Brewster to Hancock, BH). The delays of all observed source pairs are shown in this figure. The high quality of the global fit is very clear, and indeed superior to the results of the epoch reported in Paper III. The rms of the post-fit undifferenced delays range from 2.2\,ps (baseline Brewster--Hancock observing source 00) to 54\,ps (baseline Kitt Peak--North Liberty observing source 20). For the differenced delays, the rms of the post-fit residuals range from 0.26\,ps (Brewster--Pie Town observing the pair 20--19) to 7\,ps (Fort Davis--Mauna Kea observing the pair 20--18). The uncertainties in all observables were scaled to the rms of the post-fit residuals, arranged for each baseline and source pairs, to minimize the effect of bad data on the final astrometric results (as also done in Paper III).

\subsection{Proper motions between 2000 and 2010 epochs}
\label{ProperMotionSec}

\begin{figure}[th!]
\centering
\includegraphics[width=9cm]{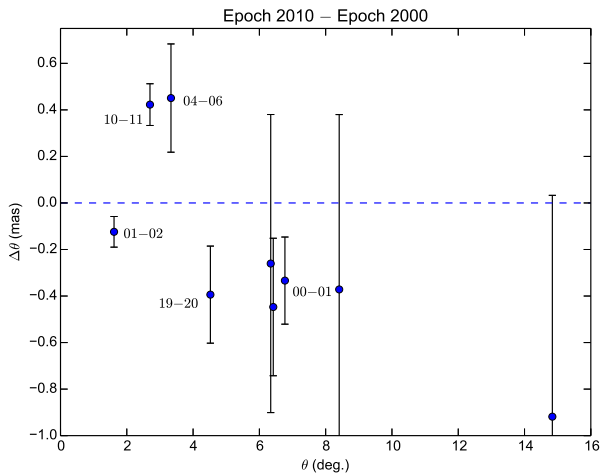}
\caption{Differences in source separations between the years 2000 and 2010, among all the source pairs commonly observed in the two epochs. The source names for the pairs with most significant motions ($>1.5\sigma$) are also shown. 
} 
\label{Astro-allsources}
\end{figure}

From these new astrometric results, we can study how much the source cores have moved between our observations in years 2000 and 2010. We summarize the results in Table \ref{CSlistORIG}. In Fig. \ref{Astro-allsources}, we show the changes in angular separation among sources as a function of source separation. 
Since source 07 (0716+714) shows a more compact structure than source 04 (0454+844) both at 14.4 and 43.1\,GHz, we decided to use this source as absolute reference source unlike in Paper III, where we used instead source 04. In any case, the choice of a different reference source does not affect the observed differences in source separations substantially (the disagreements are well within the error bars), since any shift in the sky (due to the shift of the reference source) keeps constant the angular separations among the sources (source separations are independent of any rotation of the sky). 

The uncertainties in the source motions between epochs 2000 and 2010 were estimated using a Monte Carlo approach. We generated a set of 500 different realizations of the fit, obtained by adding random (and fixed) tropospheric delays, ionospheric delays, and antenna-position shifts. The tropospheric delays at each station were changed following a Gaussian distribution with a variance of 0.2\,ns (i.e., similar to the uncertainties of the tropospheric delays fitted by UVPAP). The ionospheric delays were modified by inserting random variations in the total electron content (TEC) above each station, following a Gaussian distribution of variance 0.1\,TECUs (i.e., the expected unmodelled contribution after our astrometry fit; see \citealt{PaperIII}). We also added random changes to the antenna positions, with a Gaussian distribution of 1\,cm variance in each of the three geocentric coordinate axis. From each Monte Carlo iteration, the motions of all pairs of sources between epochs 2000 and 2010 was computed. The uncertainties in the motions were then obtained from their standard deviation over all the Monte Carlo iterations. The contributions to the error budget related to other (non-atmospheric) effects, such as station clocks or UT1$-$UTC, are much smaller than those included in the Monte Carlo analysis, and were already taken into account in the estimate of the position uncertainties made by UVPAP, which are based on the post-fit covariance matrix. These (small) extra uncertainties were added in quadrature to those from the Monte Carlo analysis.

\begin{table}
\caption{Results for the source pairs at the 15\,GHz band: the displacements indicate the change in separation among source cores between the two epochs. }
\label{CSlistORIG}
\begin{center}
\begin{tabular}{lc}\hline\hline
Pair & 2000$-$2010 displacement (mas) \\\hline
01 - 00 &    $-$0.334 $\pm$     0.190 \\
01 - 02 &    $-$0.123 $\pm$     0.065 \\
04 - 06 &    $+$0.451 $\pm$     0.230 \\
08 - 07 &    $-$0.440 $\pm$     0.300 \\
11 - 10 &     $+$0.423 $\pm$     0.090 \\
11 - 18 &    $-$0.920 $\pm$     0.970 \\
18 - 17 &    $-$0.371 $\pm$     0.750 \\
18 - 20 &    $-$0.394 $\pm$     0.210 \\
19 - 20 &    $-$0.260 $\pm$     0.640 \\
\end{tabular}
\end{center}
\end{table}

On average, and in absolute value, the source pairs have changed their separations by 0.26$\pm$0.20\,mas (compatible with zero), although there is a pair of sources, 11--10, for which a non-zero proper motion is detected at 4.7$\sigma$. 
The proper motions of the 14.4\,GHz cores, averaged over a decade, are thus in the range 0$-$100\,$\mu$as\,yr$^{-1}$.  
A comparison of the images of all sources in years 2000 and 2010 (see Fig. \ref{sources15}) indicates that a substantial evolution in the source structures has taken place over a decade in some of them, with differences in the contour locations (with respect to the position of the peak intensity) of the order of a large fraction of  a milliarcsecond (this is specially true for source 11). 
These results suggest that a small fraction of the jet cores (at least, 11) whose locations are believed to be relatively stable, compared to those of optically-thin features \citep{Blandford}, do change after a few years their absolute positions in the sky at levels higher than the astrometry precision of current and future AGN-based inertial reference frames. 

\cite{Titov} have reported on proper motions for a large sample of radio-bright AGN, from global geodetic and astrometric VLBI observations spanning several decades. Indeed, all the S5 polar-cap-sample sources, but 02, have peculiar motions reported by \cite{Titov}. The average peculiar motion for the S5 sources at 8\,GHz, according to \cite{Titov}, is 99$\pm$65\,$\mu$as\,yr$^{-1}$, which is of the order of the peculiar motions that we report for the same sources at 15\,GHz (i.e., 0$-$100\,$\mu$as\,yr$^{-1}$).

There are other cases of AGNs where systematic motions have been found in their jet cores from intensive VLBI campaigns in phase-referencing mode, either at several frequencies \citep[e.g., ][]{M81} or at a single frequency \citep[e.g., ][]{Bartel2012}. In some sources, the core motions appear to be periodic \citep[likely due to jet precession, e.g., ][]{M81,Nadia,Lob3C} and could be the effect, for instance, of either binary central engines or large-scale hydrodynamical instabilities \citep[e.g., ][]{Perucho}. More random jitterings found in other jet cores \citep[e.g., ][]{Bartel2012} could be due to a randomly-changing activity in the central engine. The peculiar core motions reported in all these works reach values of up to several tens of $\mu$as\,yr$^{-1}$, which are similar to the proper motions reported here for the S5 polar cap sample. These results challenge the picture of an astrometrically-stable AGN jet core. 

\section{Results at 43.1\,GHz}
\label{results43}

\begin{figure*}[th!]
\centering
\includegraphics[width=12cm]{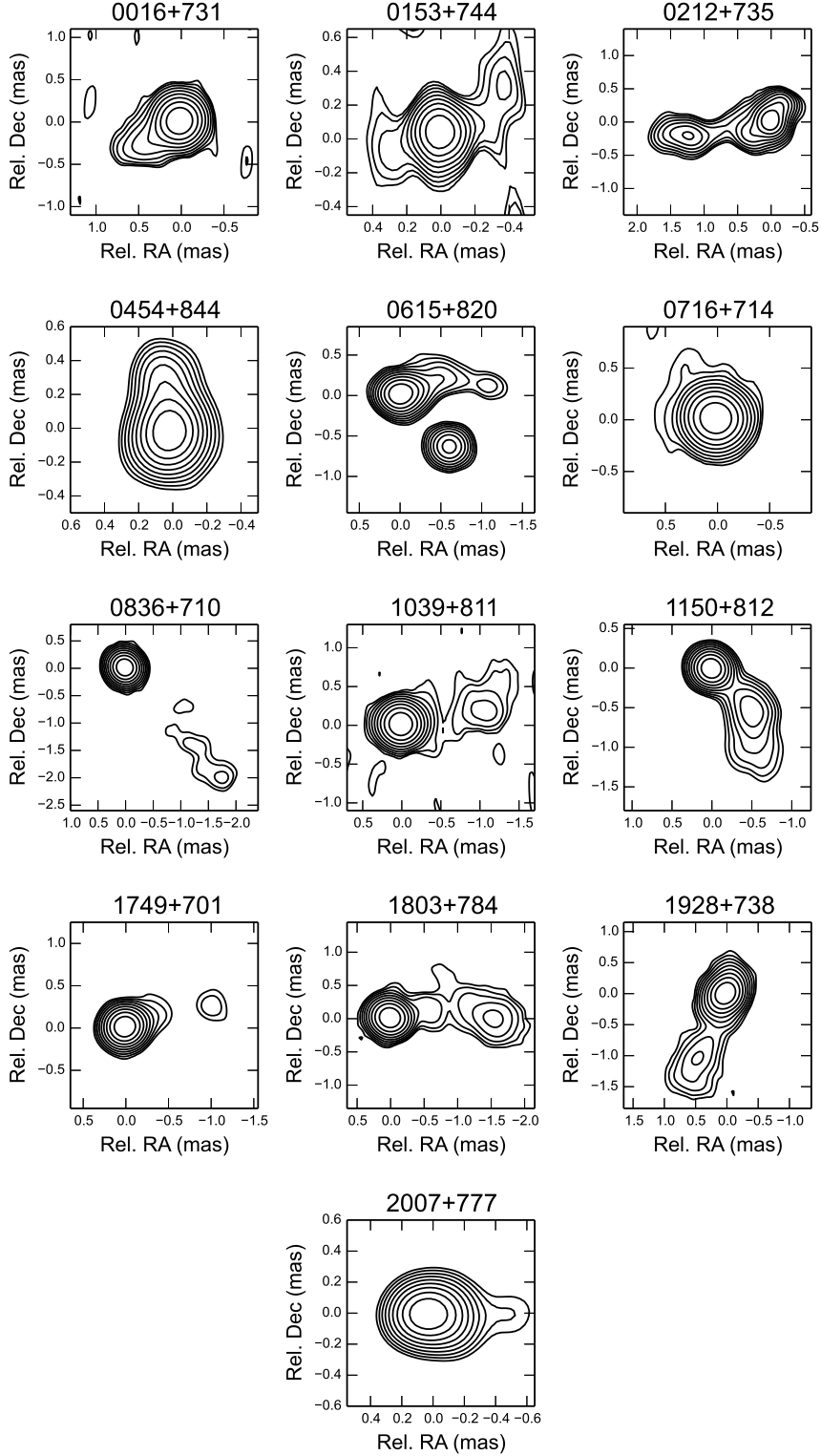}
\caption{Images of the S5 polar cap sample sources at 43.1\,GHz. The sources have been shifted with their intensity peaks at the coordinate origins. The contours are spaced logarithmically, from 0.5\% to 99\% of the source intensity peaks (for sources 00, 07, 08, 10, 17, and 18), from 2\% to 99\% (for sources 02, 06, 11, 19, and 20), and from 10\% to 99\% (for sources 01 and 04). The restoring beams are 0.3$\times$0.3\,mas FWHM.}
\label{sources43}
\end{figure*}

In Fig. \ref{sources43}, we show the structures of all sources at 43.1\,GHz, as observed in year 2010. Similar to Fig. \ref{sources15}, all images in Fig. \ref{sources43} have been shifted to set their intensity peaks at the coordinate origin of each image. The ten contours shown are spaced logarithmically, from 0.5\% to 99\% of the source intensity peaks (for sources 00, 07, 08, 10 and 18), from 2\% to 99\% (for sources 02, 06, 11 and 17) and from 10\% to 99\% (for sources 01, 04 and 20). The restoring beam in all cases is set to 0.3$\times$0.3\,mas of FWHM. The intensity peaks of all the maps at 14.4\,GHz and 43.1\,GHz are given in Table \ref{SOURCELIST2}.

From all images shown, there are a few cases that deserve additional comments. The jet extension of source 04 (0454+844) at 14.4\,GHz is seen towards the south (see Fig.\,\ref{sources15}), while the jet extension at 43.1\,GHz is seen apparently the other way around, with the brightest feature (i.e., the core) at south. The north extension at 43.1\,GHz could be due, for instance, to a jet feature (hot spot) propagating downstream from the jet base. If the feature is approaching the 43.1\,GHz core (but it is still in the self-absorbed region), it could be seen, morphologically, as a false jet-like extension towards north (or as a false counter-jet in the direction to the jet base). The core-shift of 04 (Sect. \ref{cshiftSec}) confirms this interpretation.

Another source worth mentioning is 06 (0615+820). At 43.1\,GHz, it shows two cores, one at north-east (NE) and the other at south-west (SW). The NE core shows a jet extension in the east-west direction, whereas the SW core seems unresolved. This geometry is quite different of what it is guessed from the 14.4\,GHz image alone (Fig. \ref{sources15}), where the two cores are blended in an apparent jet-like structure in the north-south direction. In Sect. \ref{SPIXSec}, we discuss on the frequency-dependent brightness distribution of this source.

\section{Comparison between 14.4\,GHz and 43.1\,GHz}
\label{results1543}

\subsection{Core shifts}
\label{cshiftSec}

In the original observational discovery  of the core-shift effect in the pair 1038+528\,A,B, the shift was of 0.7 mas and it was determined rigorously with precision differential astrometry at the micro-arcsecond level \citep{Marcaide89,Marcaide84}. 
\cite{Kovalev}, using the method of registration of optically thin components in the source structure, reported a statistical study of the core-shift effect, between 2.3\,GHz and 8.4\,GHz, in a sample of 29 AGN. These authors found a median in the core-shift of 0.44\,mas (with extreme values as large as 1.4\,mas). Assuming particle-field energy equipartition in a smooth compact jet, the core-shifts between 14.4\,GHz and 43.1\,GHz can be related to those between 8.4\,GHz and 2.3\,GHz \citep[see Eq.\,11 in][]{Lobanov}. Under these assumptions, the shifts between 14.4\,GHz and 43.1\,GHz would be 0.15 times those between 2.3\,GHz and 8.4\,GHz. Hence, if the statistical study by \cite{Kovalev} can be extrapolated to the whole S5 polar cap sample, we would expect to find typical 14.4\,GHz/43.1\,GHz core shifts of 0.06\,mas, with extreme values of about 0.21\,mas.

\begin{figure*}[th!]
\centering
\includegraphics[width=15cm]{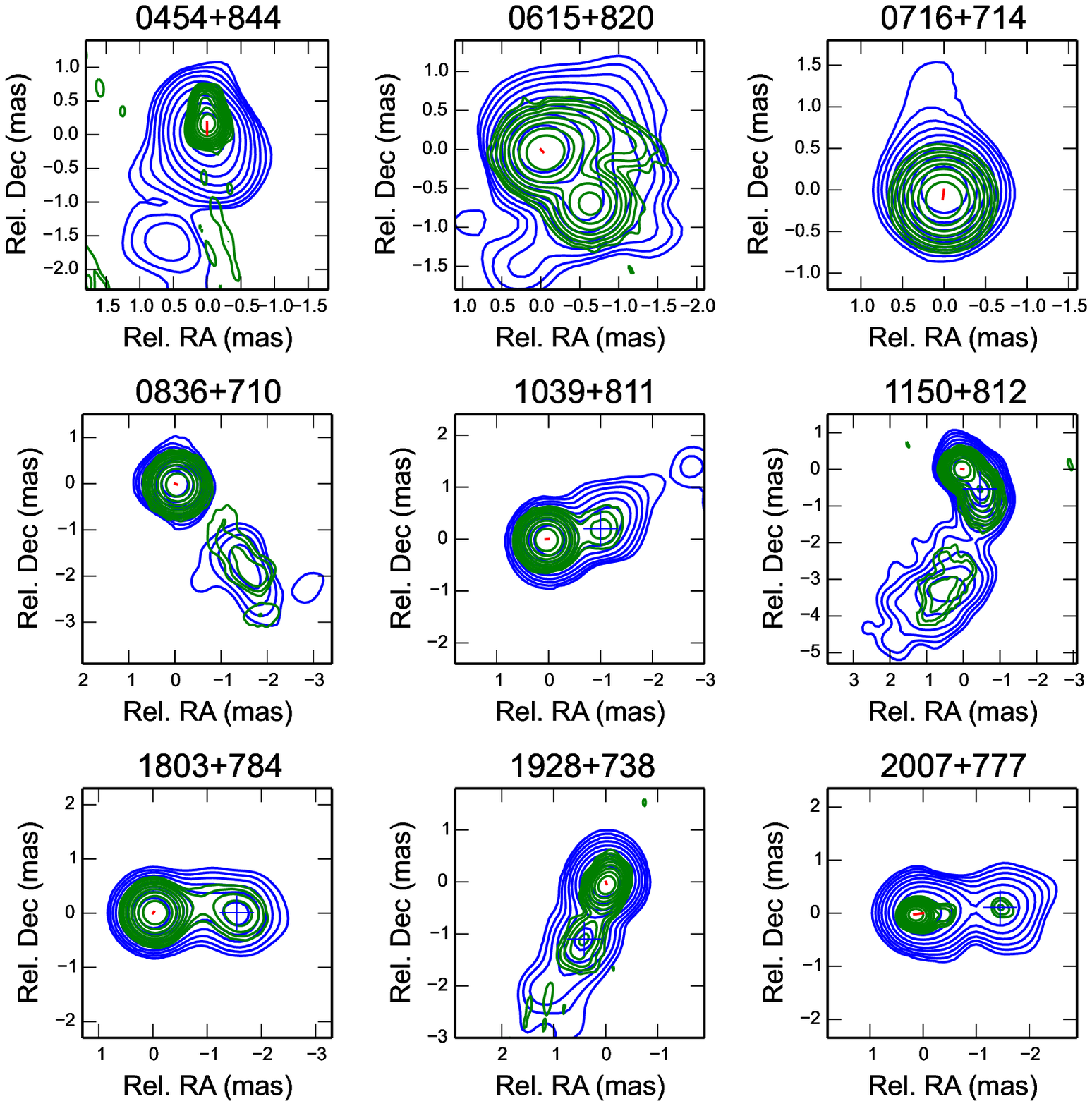}
\caption{Sources with a successful phase-transfer calibration. Contours at 14.4\,GHz are shown in blue; at 43.1\,GHz, in green. Notice the short red lines close to the image peaks, which indicate the direction and magnitude of the core shifts. The FWHM of the restoring beam in all images is 0.6$\times$0.6\,mas. The optically-thin components used as astrometry references in the SFPR analysis (Sect. \ref{FFSCal}) are indicated with crosses. The uncertainties in the core shifts (Table \ref{SOURCELIST2}) are not shown in this figure for clarity.}
\label{CSFig}
\end{figure*}

We list the measured core shifts in Table \ref{SOURCELIST2}. The uncertainties are computed from a Monte Carlo analysis, by letting the peak position of each source (and its SFPR calibrator) vary following a random Gaussian distribution, of shape equal to the restoring beam (using natural weighting) divided by the S/N of the SFPR images. The positions of the compact optically-thin components used as astrometry references (see Sect.\,\ref{FFSCal} and Fig.\,\ref{CSFig}) are also changed in the Monte Carlo analysis, following Gaussian distributions of size equal to the restoring beams, divided by the S/N of the component peaks at 43.1\,GHz.

Our core shifts are, within uncertainties, of the order of the values expected from the aforementioned extrapolation of the typical shifts found by \cite{Kovalev} in their statistical analysis of AGN. In Fig. \ref{CSFig}, we show (blue contours) the 14.4\,GHz images (shifted with their peaks at the coordinate origin) and (in green contours) the 43.1\,GHz images, shifted according to the phase-transfer astrometry results. The restoring beams of all images have been set to 0.6$\times$0.6\,mas FWHM (i.e., a compromise beam for 14.4 and 43.1\,GHz). The magnitude (and direction) of the core shifts is shown as red lines.

\begin{table*}
\caption{Results for individual sources: map peak intensities at the epochs 2000 (only 15\,GHz) and 2010 (14.4 and 43.1\,GHz) and shift of those peak intensities (usually associated to the cores, and hence called core-shifts) between 14.4 and 43.1\,GHz at epoch 2010. All shifts are re-referenced to optically-thin jet components (see Fig. \ref{CSFig}). Source 06 is a special case (see text). }
\label{SOURCELIST2}
\begin{center}
\begin{tabular}{cccccccccc}\hline\hline
Source name & Alias & PR-Calib & Peak (2000)  & \multicolumn{2}{c}{Peaks (2010)}  & \multicolumn{3}{c}{Core shift} & Ref.\\
            &       &         & (Jy/beam) & \multicolumn{2}{c}{(Jy/beam)} & RA & Dec & Module &   \\
            &       &         & 15\,GHz &  14.4\,GHz & 43.1\,GHz &   ($\mu$as)   & ($\mu$as) & ($\mu$as) & \\\hline       
B0016+731 & 00  & $-$         & 0.73 & 0.98 & 0.53 &         $-$     &  $-$          &  $-$   &  $-$ \\
B0153+744 & 01  & $-$         & 0.19 & 0.08 & 0.01 &         $-$     &  $-$           & $-$   & $-$  \\
B0212+735 & 02  & $-$         & 1.69 & 1.69 & 0.48 &         $-$     &  $-$           & $-$   & $-$  \\
B0454+844 & 04  & 06          & 0.17 & 0.17 & 0.07 &  2 $\pm$ 138  &  181  $\pm$ 74  &   226  $\pm$ 78 & 10\\
B0615+820 & 06  & 04, 07      & 0.27 & 0.38 & 0.14 &  $-$32 $\pm$ 95  &  $-$35  $\pm$ 43  &   100  $\pm$ 56 & 10\\
B0716+714 & 07  & 06, 08      & 1.02 & 2.04 & 1.62 &  14 $\pm$ 55  &  $-$108  $\pm$ 26  &   121  $\pm$ 30  &  10\\
B0836+710 & 08  & 07, 10      & 1.42 & 1.81 & 1.41 &  $-$44 $\pm$ 53  &  $-$16  $\pm$ 25  &   66  $\pm$ 35 &  10\\ 
B1039+811 & 10  & 08     & 0.75 & 0.50 & 0.44 &  57 $\pm$ 12  &  $-$4  $\pm$ 13  &   58  $\pm$ 12  & 10\\
B1150+812 & 11  & $-$          & 0.55 & 0.43 & 0.16 &  54 $\pm$ 2  &  9  $\pm$ 2  &   55  $\pm$ 2  & 11\\
B1749+701 & 17  & $-$         & 0.31 & 0.47 & 0.22 & $-$ & $-$ & $-$ & $-$ \\
B1803+784 & 18  & 19, 20      & 1.79 & 1.70 & 0.84 &  $-$20 $\pm$ 4  &  24  $\pm$ 4  &   31  $\pm$ 3 &  18\\
B1928+738 & 19  & 18, 20      & 1.53 & 3.24 & 1.39 &  $-$15 $\pm$ 4  &  $-$36  $\pm$ 5  &   39  $\pm$ 5 &  19\\
B2007+777 & 20  & 18, 19      & 0.95 & 0.60 & 0.22 &  168 $\pm$ 8  &  $-$23  $\pm$ 9  &   170  $\pm$ 8 & 20 \\

\end{tabular}
\end{center}
{Notes:}~ {\em PR-Calib} are the aliases of the sources used as SFPR calibrators for each source (see Sect. \ref{FFSCal}). {\em Core shift} is the shift between the intensity peaks of the images at different frequencies. We notice, though, that the intensity peak may not correspond to the true core of the AGN jet in some cases (see Sect. \ref{SPIXSec}). The intensity peaks correspond to restoring beams of 0.6$\times$0.6\,mas at all frequencies. The VLBI uncertainties in the absolute flux-density calibration are typically 5$-$10\%. {\em Ref} is the source with an optically-thin jet feature used as a position reference for the core shift (see text).
\end{table*}

As can be seen in Fig. \ref{CSFig}, for sources 04, 08, 10, 11 and 20, the core-shifts are roughly aligned with the jet direction. This is an expected result, since the core shift (due to synchrotron self-absorption) shall be aligned with the highest magnetic-field and/or particle-density gradient, which is given in the direction longitudinal to jet \citep[e.g., ][]{Lobanov}. However, for sources 06, 07, 18 and 19, we find hints of core shifts misaligned to the prominent jet directions. 
Regarding source 06, it is difficult to tell whether the shift is also parallel to the innermost part of the jet at 43.1\,GHz, given the complex source structure found at that frequency.

\subsection{Spectral-index images}
\label{SPIXSec}

\begin{figure*}[th!]
\centering
\includegraphics[width=12cm]{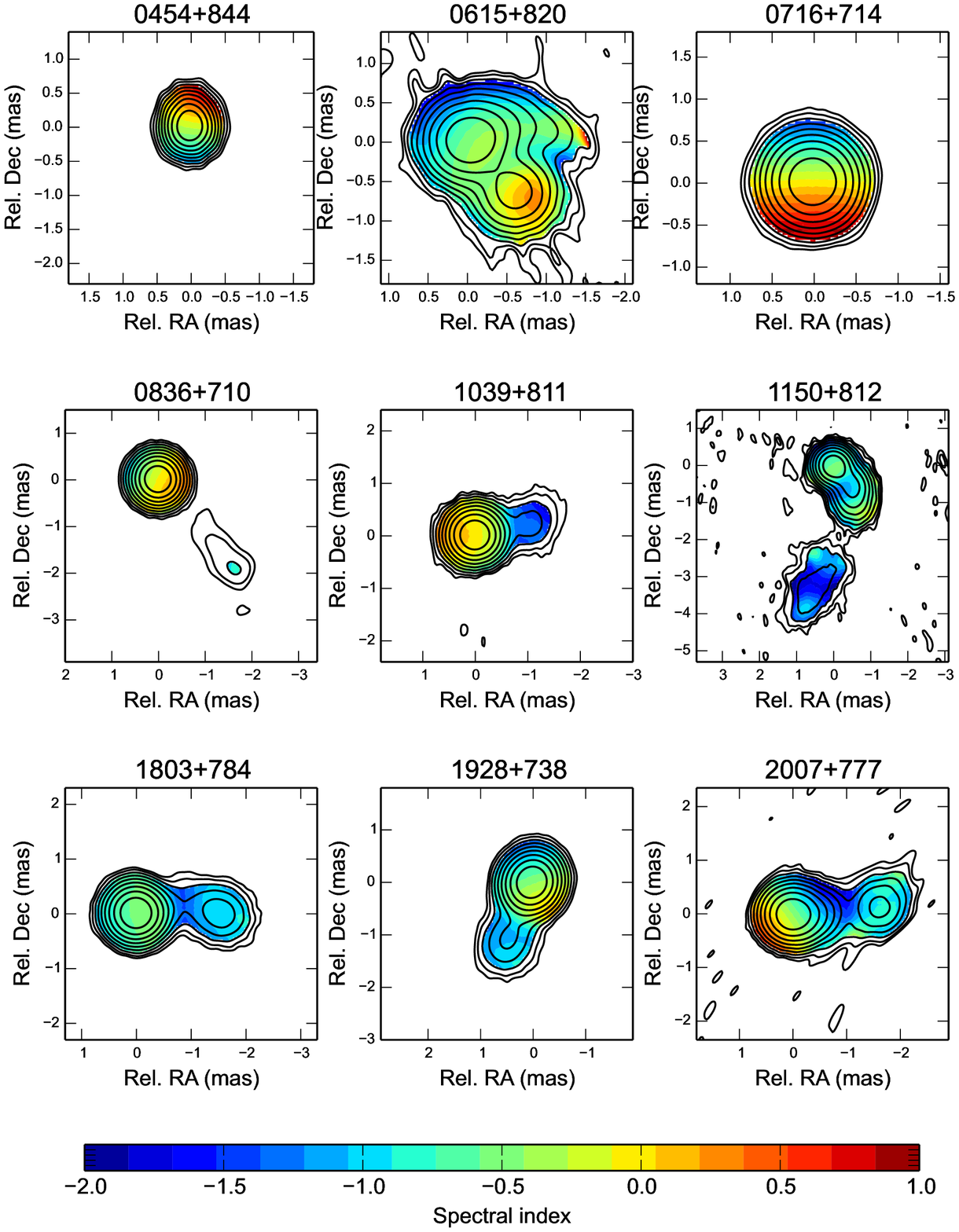}
\caption{Spectral-index distribution of a subset of the S5 sources (see text). The contours correspond to the 43.1\,GHz images, convolved with a beam of 0.6$\times$0.6\,mas FWHM. We notice that the effects of the core-shift uncertainties (Table \ref{SOURCELIST2}) are not shown in this figure.}
\label{SPIX}
\end{figure*}

We show in Fig. \ref{SPIX} the spectral-index distribution (i.e., $\alpha$, being the flux density $\propto \nu^\alpha$) for the sources where a successful phase-transfer calibration could be performed. For the spectral-index computation, we have used a compromise convolving beam of 0.6$\times$0.6\,mas FWHM for the images at 14.4\,GHz and 43.1\,GHz. In Fig. \ref{SPIX}, we see that the jet cores have nearly flat (or even inverted) spectra, with the spectral index, $\alpha > 0$. This is due to synchrotron self-absorption effects in the core region. The jet extensions are, however, optically thin, with $\alpha < 0$. Similar distributions of spectral index have been found in many other AGN jets \citep[e.g., ][]{Marcaide84, Kovalev} and are well understood in terms of the standard jet model. We notice that, since the 14.4\,GHz images have been over-resolved, there can be artifacts in the spectral-index distributions, especially in the regions close to the lowest contours.

A peculiar case is source 06. The hardest spectrum (i.e., highest spectral index) is found on the SW component. This would be the ``spectral core'' of the source. 
However, the intensity peak at both 14.4\,GHz and 43.1\,GHz (i.e., what we could call the ``morphological core'') is located at the NE component. 
In any case, a clear result is that the SW emission does show clear signs synchrotron self-absorption (being thus more likely close to an AGN central engine) and is much misaligned to the 43.1\,GHz jet extension propagating towards west from the NE component. A possible explanation for this morphology could be that the true core of the jet is the SW component, with the jet propagating towards NE. The hot spot at NE could be due, for instance, to a strong interaction region of the jet with its surrounding medium, which would break or re-direct the jet toward the west (causing the east-west jet extension seen in the 43.1\,GHz image of the NE component). This interpretation, though, would be insufficient to explain another intriguing morphological feature in this source: there is a hint of ring-like structure in the image at 14.4\,GHz, with a diameter of $\sim$1\,mas. This structure resembles the image at 5\,GHz reported by \cite{Dodson} from space-VLBI observations. A possible explanation for such a structure might be a mas-size gravitational lens, although a more complete multi-frequency (and full-polarization) analysis should be performed to confirm this possibility. Another possibility would be a strongly bent jet oriented to the line of sight, as it is seen in, for example, PKS\,2136$+$141 \citep{Savolainen}, although the fact that the ring structure in 06 is unbroken would be difficult to explain in that scenario.

\begin{figure}[th!]
\centering
\includegraphics[width=10cm]{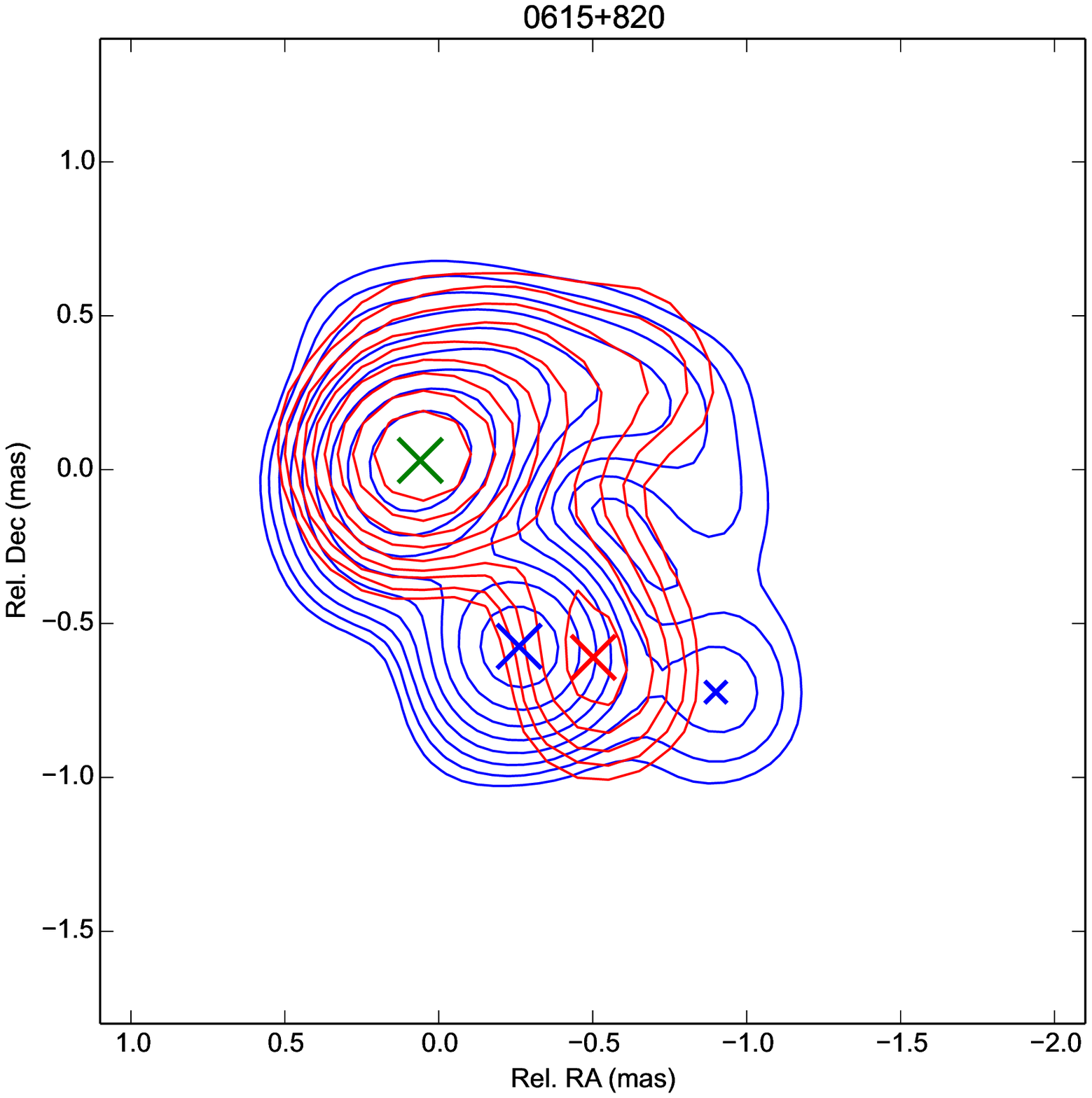}
\caption{Images of source 06 at the 15\,GHz band in the year 2000 (blue contours) and 2010 (red contours). The restoring beam is 0.4$\times$0.4\,mas and the ten contours are spaced logarithmically from 5\% to 99\% of the peak intensities (0.15 and 0.27\,Jy/beam for year 2000 and 2010, respectively). The red and blue crosses indicate the position of the SW component in 2010 and 2000 (respectively). The green cross shows the location of the NE component (set equal in both epochs). A second peak at west of SW in the year 2000 is also shown by a smaller blue cross.}
\label{04-2000}
\end{figure}

An alternative explanation could be that NE and SW are the cores of different AGN, so that 0615$+$820 would be a binary massive black hole. A precise astrometric follow-up between NE and SW at high frequencies, together with simultaneous observations at lower frequencies (to study the evolution of the spectral-index distribution) would be required to confirm this possibility. In Fig. \ref{04-2000}, we show the over-resolved images of source 06 at the 15\,GHz band for year 2000 (blue contours) and 2010 (red contours), using the NE component as position reference. We indicate with crosses (of the same contour colors) the location of the SW component at each epoch. An intriguing shift is seen between the two epochs, which might be caused by an orbital motion of SW in a binary massive black hole. In addition, a third weaker component can be seen to the west of SW in year 2000. A deeper analysis of the binary black hole scenario (and other alternative explanations) for source 06, using all available VLBI data of this source, will be published elsewhere.

\section{Conclusions}
\label{conclusions}

We report on quasi-simultaneous 14.4\,GHz and 43.1\,GHz VLBA observations of the S5 polar cap sample, performed in December 2010 in phase-referencing mode, using the fast-frequency-switching (FFS) capabilities of the VLBA, and compare them to earlier results at 15\,GHz band. We have performed a high-precision (differential phase-delay) analysis at 14.4\,GHz, solving for all the $2\pi$ phase ambiguities as in Paper III. Between the years 2000 and 2010, we find a $4.7\sigma$ proper motion of 42$\pm$9\,$\mu$as\,yr$^{-1}$ between the jet cores of sources 10 and 11. For the rest of source pairs, the separations did not change above 2\,$\sigma$. 

We have performed an SFPR calibration, from 14.4\,GHz to 43.1\,GHz, to determine the core shifts. Only nine of the thirteen sources could be imaged with this technique. We find typical core-shifts of 0.05$-$0.2\,mas. We have constructed robust spectral-index images of these nine sources. The spectral-index distributions follow the well-known steepening of the spectrum at the jet extensions, from an either flat- or inverted-spectrum regions associated to jet cores. 

There is one source, 0615$+$820, that shows a remarkable double structure at 43.1\,GHz (two components, one at northeast, NE, and one at southwest, SW), having one of them, NE, a prominent jet extension roughly perpendicular to the NE$-$SW direction. 
Possible explanations for this intriguing source structure could be either a strong jet bending at parsec scales from the AGN central engine (due to interaction with the ISM), a gravitational lens with mas scale, or a binary massive black hole. The relative astrometry between NE and SW at 15\,GHz, using image over-resolution, shows a clear position drift of SW with respect to NE between years 2000 and 2010, thus supporting the third possibility (binary black hole). A deeper analysis of the results on this source, using all the available VLBI data, will be published elsewhere. Future observations at mm-wavelengths (with the Global mm-wave VLBI Array, GMVA) and at cm-wavelengths (using the RadioAstron satellite) are being planned.

\begin{acknowledgements}
IMV thanks the Alexander von Humboldt Foundation for his post-doctoral fellowship in years 2009$-$2011 (which covered a part of the research work reported here). We thank R. Dodson and M. Rioja for useful discussion. This work has been partially supported by the Spanish MINECO projects AYA2009-13036-C02-02 and AYA2012-38491-C02-01 and by the Generalitat Valenciana projects PROMETEO/2009/104 and PROMETEOII/2014/057. MPT acknowledges support by the Spanish MINECO through grants
AYA2012-38491-C02-02 and AYA2015-63939-C2-1-P, cofunded with FEDER funds.
The National Radio Astronomy Observatory is a facility of the National Science Foundation operated under cooperative agreement by Associated Universities, Inc.

\end{acknowledgements}

\end{document}